\documentclass[atmp]{ipart_v1}

\Vol{00}
\Issue{0}
\Year{2017}
\firstpage{001}

\usepackage{t1enc}
\usepackage[latin1]{inputenc}
\usepackage[english]{babel}

\usepackage{amsthm}
\usepackage{yfonts}

\usepackage{bbm}
\usepackage{bm}
\usepackage{mathrsfs}

\usepackage{graphicx}
\usepackage{epstopdf}

\numberwithin{equation}{section}

\begin{document}


\title[Stationary solutions of cubic nonlinear {Schr{{\"o}}dinger} equations]{New stationary solutions of the cubic nonlinear {Schr{{\"o}}dinger} equations \\ for Bose-Einstein condensates}


\author[Q. D. Katatbeh and D. M. Christodoulou]{Qutaibeh D. Katatbeh and Dimitris M. Christodoulou}

\begin{abstract}
We have previously formulated a simple criterion for deducing the intervals of oscillations in the solutions of second-order linear homogeneous differential equations. In this work, we extend analytically the same criterion to the cubic nonlinear {Schr{{\"o}}dinger} equations that describe Bose-Einstein condensates. With this criterion
guiding the search for solutions, we classify all types of solutions and we find new stationary solutions in the free-particle cases that were not noticed previously because of limited coverage in the adopted boundary conditions.
The new solutions are produced by the nonlinear terms of the differential equations
and they continue to exist when various external potentials are also incorporated. Surprisingly, these solutions appear when the
nonlinearities are small.
\end{abstract}

\maketitle

\section{Introduction}\label{intro}

The ordinary second-order linear homogeneous differential equations 
of mathematical physics 
\begin{equation}
y'' + b(x) y' + c(x) y = 0,
\label{general_form}
\end{equation}
can all be transformed to the canonical form
\begin{equation}
u'' + q(x) u = 0,
\label{canonical_form}
\end{equation}
where the primes denote derivatives with respect to the independent variable $x$,
\begin{equation}
q = -\frac{1}{4}\left(b^2 + 2b' - 4c\right), 
\label{theq}
\end{equation}
and $y(x) = u(x)\exp(-\frac{1}{2}\int{b(x) dx})$
\cite{abr72,whi20,har64}.
The canonical form~(\ref{canonical_form}) condenses 
the coefficients of eq.~(\ref{general_form}) into $q(x)$ and `oscillation theory'
focuses on this coefficient in order to derive the oscillatory properties 
of the solutions of eq.~(\ref{general_form}) (see the reviews in \cite{agr02,won68} 
and references therein).

In recent work \cite{chr16a,kat16}, we showed that the canonical form
is degenerate in the sense that different equations of the form~(\ref{general_form}) can be transformed to the same canonical form.
This is evident from eq.~(\ref{theq}) in which $q(x)$ is the result of combining two unrelated functions $b(x)$ and $c(x)$. Furthermore, the derivative $b'(x)$
in eq.~(\ref{theq}) sometimes acts as damping (when $b' > 0$) and other times enhances oscillations in the solutions (when $b' < 0$). We worked around these ambiguities by transforming eq.~(\ref{canonical_form}) to a form with constant damping (eq.~(\ref{general_form}) with $b=$ constant), and then we transformed
again to a new canonical form in which the constant term $b$ acted
unambiguously as damping opposing oscillatory tendencies in the solutions,
just as it does in the well understood case of the damped harmonic oscillator
(eq.~(\ref{general_form}) with $b, c=$ constant).
This procedure was very successful in deducing the precise intervals of oscillations in the solutions of the general form~(\ref{general_form}).

In the last step of the procedure, a generalized Euler transformation of the independent variable $x$ was used \cite{chr16a,kat16}:
\begin{equation}
x = c_1 + c_2\exp(kt),
\label{euler_tran}
\end{equation}
where $c_1$, $c_2$, and $k$ are arbitrary constants, and a criterion
for the intervals of oscillations in the solutions was established:
\begin{equation}
q(x) > \frac{1}{4(x - c_1)^2}.
\label{crit_1}
\end{equation}
Only the constant $c_1$ appears in the criterion and corresponds to a `horizontal shift' of the
independent variable $x$ in eq.~(\ref{euler_tran}). 
For equations with singularities
at the origin, $c_1$ can be set to zero and then the criterion~(\ref{crit_1}) reduces to the simple form
\begin{equation}
q(x) > \frac{1}{4x^2}.
\label{crit_2}
\end{equation}
In this case, we can also choose $c_2=1$ and $k=1$ in eq.~(\ref{euler_tran}) 
and then the change of the independent variable $x$ takes the form of the 
classical Euler transformation
\begin{equation}
x = \exp(t),
\label{class_euler_tran}
\end{equation}
for which the investigation of the interval $t\in (-\infty, \ +\infty)$ in the transformed
equation corresponds to searching for oscillatory solutions in the interval 
$x\in (0, \ +\infty)$ of the original equation~(\ref{general_form}).

In this work, we extend the applicability of the criterion~(\ref{crit_2})
to cubic nonlinear {Schr{{\"o}}dinger} (CNLS) equations 
of the 1+1 type, the 2+1 type, and the 3+1 type \cite{mal13,mal14a,mal14b,mal15}. 
In this notation, the $+1$ signifies the time ($t$)
dimension whereas the first digit signifies the $N$ spatial dimensions.
These equations, sometimes referred to as the Gross-Pitaevskii equations \cite{lif80}, take the form
\begin{equation}
i\hbar\frac{\partial\Psi}{\partial t} = \left( -\frac{\hbar^2}{2m}\vec{\nabla}^2 + V(\vec{x}) + g|\Psi|^2 \right)\Psi \ ,
\label{eq_main}
\end{equation}
where $m$ is the mass of the particle described by the time-dependent wavefunction $\Psi(\vec{x}, t)$, $\vec{x}$ is the vector of the spatial coordinates in
$N$=1, 2, or 3 dimensions, $V$ is the scalar potential,
$\hbar = h/2\pi$ is the reduced Planck constant, and $g$ is the amplitude of the nonlinearity.
Eq.~(\ref{eq_main})
models repulsive ($g>0$) and attractive ($g<0$)
Bose-Einstein condensates (BECs) with a variety of confining potentials $V(\vec{x})$
that serve as spatial traps of solitary waves in many applications of current
interest. 
Mallory \& Van Gorder \cite{mal13,mal14a,mal14b,mal15} have given a detailed list of current applications
of BECs, as well as the solutions for both bright and dark solitons obtained
by using a proper set of boundary conditions. 

The advantage of using the criterion~(\ref{crit_2}) to predict the conditions
for oscillatory spatial solutions is that the results do not rely on any
adopted boundary conditions. In the first step of the procedure, we search
for trivial solutions in the equations. These are necessary (but not sufficient) in order to provide a baseline for oscillations. The CNLS equations have three trivial solutions each of which can serve as a baseline for different types of oscillatory BECs.  In the next step,
the criterion~(\ref{crit_2}) distinguishes the oscillatory solutions from other `unstable' ({\it i.e.}, nonoscillatory) solutions \cite{mal13,mal14a,mal14b,mal15} based on the adopted boundary conditions in various applications. 

It turns out that all the coordinate types of the CNLS equations
can be investigated simultaneously because the dominant nonlinear terms
have the same structures in all the attractive and all the repulsive cases, respectively.
All of these cases can be covered by the same criterion for oscillations as follows:
The inertial term of the cylindrical 2+1 type is $y'/x$ and it implies that the differential
equations contain no damping of the oscillations \cite{chr16a}. Then, the criterion~(\ref{crit_2})
for oscillatory solutions reduces to the simple inequality
\begin{equation}
c(x) > 0 ,
\label{crit_3}
\end{equation}
where $c(x)$ represents the coefficient of the $y$-term in equations that can be cast in the form~(\ref{general_form}). On the other hand, the inertial terms of the 1+1 and the 3+1
types are $0y'$ and $2y'/x$, respectively, and in both of these cases the criterion~(\ref{crit_2}) reduces to $c(x) > 1/(4x^2)$ \cite{chr16a}. The term $1/(4x^2)$
represents the low-level inertial damping that is present in the cartesian and the spherical
forms of the CNLS equations, but this term becomes negligible for $x>>1$, in which case
the 1+1/3+1 criterion quickly approaches asymptotically the inequality~(\ref{crit_3}) of the 2+1 case.

In what follows, we demonstrate the analytic procedure presented in \cite{chr16a,kat16} for Bose-Einstein free solitons. This is the first time that nonlinear differential
equations with more than one trivial solution have been investigated.
(In \cite{chr16b}, the nonlinear polytropic Lane-Emden equations that possess only one trivial solution were analyzed.)
The repulsive and the attractive CNLS equations are analyzed in Sec.~\ref{rep} and Sec.~\ref{att}, respectively. In both cases, we classify the various types of solutions expected to exist depending on the adopted boundary conditions; and we verify the results by high-accuracy numerical integrations \cite{sha97,sha99} of the 2+1 type for which the criterion~(\ref{crit_3}) is exact.\footnote{We note that additional numerical calculations for the 1+1 and 3+1 cases with and without external potentials (not shown here) also verify the same nonlinear properties of the solutions.} Finally, in Sec.~\ref{sum}, we discuss our results.

\begin{figure}[h]
\includegraphics[width=\linewidth]{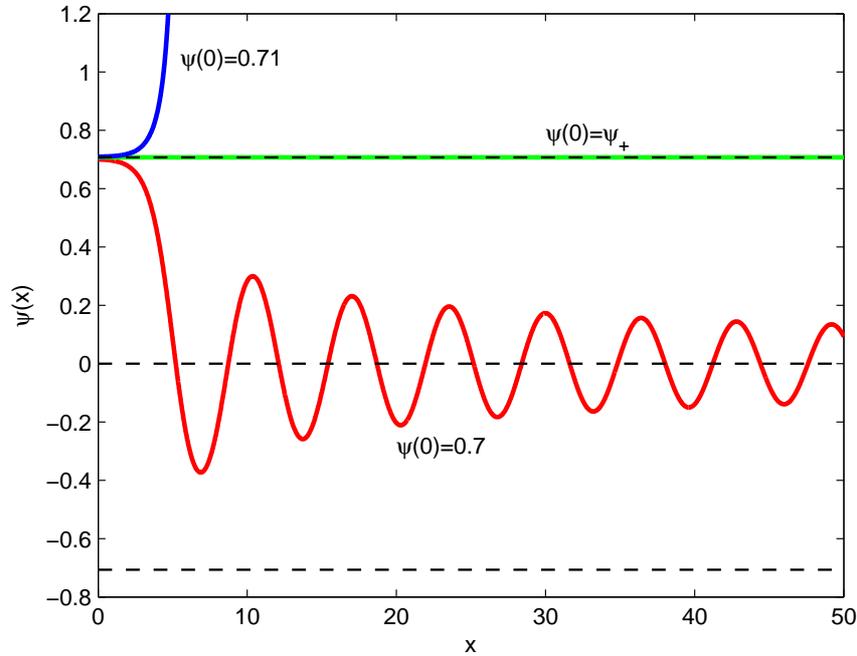}
\caption{The numerical solutions of the cylindrical eq.~(\ref{eq_r1}) 
with $N=2$, $\psi'(0)=0$ and boundary values $\psi(0)=$0.7 (red curve), $1/\sqrt{2}$
(green line), and 0.71 (blue curve). The trivial solutions are also shown as dashed lines.
}
\label{fig1}
\end{figure}

\section{Repulsive CNLS equations}\label{rep}

Following the extensive study in \cite{mal14b} (their eqs.~(1) to~(14)), we adopt the dimensionless CNLS equation
\begin{equation}
\psi'' + \frac{N-1}{x}\psi' + \left(1 - 2|\psi|^2\right)\psi = 0 , \ \ (N=1,2,3),
\label{eq_r1}
\end{equation}
to describe the spatial part of the stationary wavefunction $\psi(x)$ in the repulsive free-particle case. 
The time-dependent part of the wavefunction in eq.~(\ref{eq_main})
is assumed to have the form $\exp\left(-i\hbar t/2m\right)$, where 
$m$ is the particle mass and $\hbar = h/2\pi$ is the reduced Planck constant.
Here we do not limit the nonlinearity to a small value, so our calculations
can accomodate arbitrarily strong nonlinear properties in the solutions. In eq.~(\ref{eq_r1}), the magnitude of the nonlinear amplitude $g>0$ is effectively set by the choice of the boundary value $\psi(0)$ since $\psi(0)\propto 1/\sqrt{g}$.

Using eqs.~(\ref{crit_3}) and~(\ref{eq_r1}), we obtain the criterion for oscillatory solutions 
\begin{equation}
|\psi|^2 < \frac{1}{2} .
\label{crit_4}
\end{equation}
This inequality predicts oscillations for as long as the wavefunction remains between
$\psi_\pm = \pm 1/\sqrt{2}$. These two values are also trivial solutions of eq.~(\ref{eq_r1}),
in addition to the better known trivial solution $\psi = 0$. Since these trivial solutions
do not satisfy the criterion~(\ref{crit_4}), then oscillations can only occur about
$\psi = 0$. In this simple way, we can classify the various solutions of eq.~(\ref{eq_r1})
as follows: (a)~For boundary conditions of the form $|\psi(0)| < \psi_+$, the solutions will be oscillatory about $\psi = 0$. (b)~For $|\psi(0)| = \psi_+$, the solutions will be constant.
(c)~For $|\psi(0)| > \psi_+$, the solutions will be repelled by the nearest nonzero trivial solution and they will diverge rapidly. Such divergent solutions were noted by Mallory \& Van Gorder \cite{mal14b} for the case of a constant potential. We see here that they are not produced by the potential, instead they have their origin in the nonlinearity of eq.~(\ref{eq_r1}).

The three types of solutions are illustrated numerically in Fig.~\ref{fig1} for the following boundary conditions: $\psi'(0)=0$ and $\psi(0)=0.7$, $1/\sqrt{2}$, and 0.71. Clearly, only the trivial 
solution $\psi = 0$ attracts nearby
oscillatory solutions while the nonzero trivial solutions repel all solutions.

Divergent solutions appear for any choice of the boundary condition 
$|\psi(0)| > 1/\sqrt{2}$. This condition implies that the nonlinearity in eq.~(\ref{eq_r1}) takes relatively small values. 
This is because the chosen value for
$\psi(0)$ scales as $1/\sqrt{g}$, where $g$ is the nonlinear amplitude
of $|\psi|^2$ in the normalization given by Mallory \& Van Gorder \cite{mal13,mal14a,mal14b,mal15} for eq.~(\ref{eq_r1}).
This result is counter to intuition as it indicates that divergent solutions appear only for small perturbations (of order $g|\psi|^2$)
in the {Schr{{\"o}}dinger} equation, while strong perturbations
with $0 < |\psi(0)| < 1/\sqrt{2}$ always lead to well-behaved oscillatory solutions about $\psi = 0$ (as the red curve in Fig.~\ref{fig1}).

\begin{figure}[h]
\includegraphics[width=\linewidth]{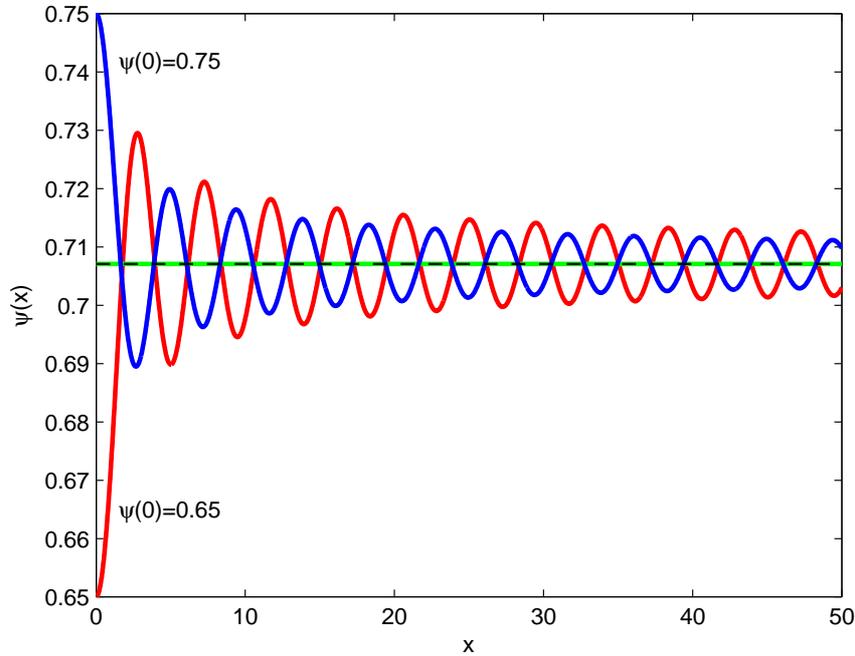}
\caption{The numerical solutions of the cylindrical eq.~(\ref{eq_a1}) 
with $N=2$, $\psi'(0)=0$ and boundary values $\psi(0)=$0.65 (red curve), $1/\sqrt{2}$
(green line), and 0.75 (blue curve). The trivial solution $\psi_+=1/\sqrt{2}$ is also marked by a dashed line.
}
\label{fig2}
\end{figure}

\section{Attractive CNLS equations}\label{att}

Following \cite{mal14b} again (their eqs.~(1) to~(14)), we adopt the dimensionless CNLS equation
\begin{equation}
\psi'' + \frac{N-1}{x}\psi' + \left(2|\psi|^2 - 1\right)\psi = 0 , \ \ (N=1,2,3),
\label{eq_a1}
\end{equation}
to describe the spatial part of the stationary wavefunction $\psi(x)$ in the attractive free-particle case. 
The time-dependent part of the wavefunction in eq.~(\ref{eq_main})
is assumed to have the form $\exp\left(+i\hbar t/2m\right)$, where again
$m$ is the particle mass and $\hbar = h/2\pi$ is the reduced Planck constant. In eq.~(\ref{eq_a1}), the magnitude $|g|$ of the nonlinear amplitude $g<0$ is effectively set by the choice of the boundary value $\psi(0)$ since $\psi(0)\propto 1/\sqrt{|g|}$.

Using eqs.~(\ref{crit_3}) and~(\ref{eq_a1}), we obtain the criterion for oscillatory solutions 
\begin{equation}
|\psi|^2 > \frac{1}{2} .
\label{crit_5}
\end{equation}
This inequality predicts oscillations for as long as the wavefunction manages to repeatedly cross outside the interval $(\psi_-, \psi_+)$ where again
$\psi_\pm = \pm 1/\sqrt{2}$. These two values are also trivial solutions of eq.~(\ref{eq_a1}), in addition to $\psi = 0$. In this case, these two trivial solutions fail marginally to 
satisfy the criterion~(\ref{crit_5}) whereas $\psi = 0$ fails completely and it will repel all solutions.
It would then appear
that oscillatory solutions can occur only around $\psi = \psi_\pm$ for those wavefunctions that manage
to satisfy the inequality $|\psi(x)|>\psi_+$ at some radii $x$.  

There exists however one case where the oscillations will develop about one or the other nonzero trivial solution after
a rather complicated behavior that involves also the repulsive trivial solution $\psi = 0$.
It turns out that
the attractive problem is not the exact inverse of the repulsive problem analyzed in Sec.~\ref{rep} because of the absence of divergent solutions and the existence of new solutions in the case of boundary
conditions with $|\psi(0)|>>1$. Obeying the criterion~(\ref{crit_5}), such solutions will
oscillate about $\psi = \psi_+$ or $\psi = \psi_-$, but in the process they can overshoot
the repulsive trivial solution $\psi = 0$ and intersect it several times. Their behavior will be
determined by the choice of $\psi(0)$ and by the fact that $\psi = 0$ works to repel
all solutions; so the new solutions are forbidden from decaying asymptotically on to $\psi=0$.

\begin{figure}[h]
\includegraphics[width=\linewidth]{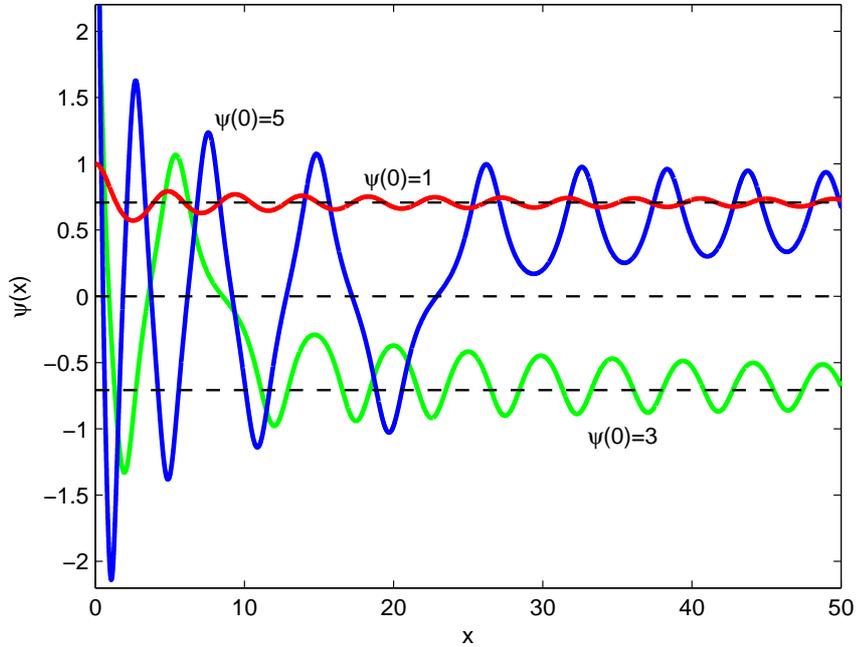}
\caption{The numerical solutions of the cylindrical eq.~(\ref{eq_a1}) 
with $N=2$, $\psi'(0)=0$ and boundary values $\psi(0)=$1 (red curve), 3 (green curve), and 5 (blue curve). The three trivial solutions are also shown as dashed lines.
}
\label{fig3}
\end{figure}

We can now classify the various solutions of eq.~(\ref{eq_a1}) 
as follows: (a)~For $|\psi(0)| = \psi_+$, the solutions will be constant.
(b)~For boundary conditions of the form $|\psi(0)|\approx 1$, the solutions will be oscillatory about $\psi = \psi_+$ or $\psi = \psi_-$. (c)~In cases where $|\psi(0)| < \psi_+$,
the solutions will be repelled by $\psi = 0$ and in the process they will have to cross
one of the nonzero trivial solutions, thereby becoming oscillatory according to eq.~(\ref{crit_5}).
(d)~For $|\psi(0)| >> \psi_+$, the solutions will still oscillate about one or the other nonzero trivial solution but in a more complicated fashion. Such solutions cannot be
discovered by examining the density $|\psi(x)|^2$ of the BEC because then the
complexities of the underlying radial ($x$) solution are lost and the repulsion
of the $\psi=0$ trivial solution is no longer visible. As in the repulsive case of Sec.~\ref{rep}, 
the normal and the exotic features of the bright solutions are caused by the
nonlinearity of eq.~(\ref{eq_a1}) and they are present in cases where an external
potential is introduced.

The first three types of solutions are illustrated numerically in Fig.~\ref{fig2} for the following boundary conditions: $\psi'(0)=0$ and $\psi(0)=1/\sqrt{2}$, 0.75, and 0.65.
The solutions of cases (b) and (c) in the classification above are clearly attracted by the $\psi=\psi_+$ trivial solution
and are forced to oscillate about it. The fourth type of (exotic) solutions are illustrated
numerically in Fig.~\ref{fig3} for $\psi'(0)=0$ and~$\psi(0)=3$ and 5 along with the normal oscillatory solution for $\psi(0)=1$. 

The exotic solutions cross the repulsive trivial solution $\psi=0$
several times but they are forbidden from settling on to it, so after a few cycles they are attracted to one of the nonzero trivial solutions. We have no way of telling which trivial solution they will be attracted to, this choice depends on the amplitude of the
decaying oscillations at internediate values of $x$. But the transition occurs always at an inflection point that develops on the $\psi=0$ line (Fig.~\ref{fig3}). 
When the inflection occurs, the wavelength becomes a lot longer (by factors of order $\sim$2) and this change in wavelength
distinguishes these solitons from all other bright solitons. For this reason,
such solitons may be easily identifiable in experiments creating actual BECs.

The exotic solutions appear for $|\psi(0)| >> 1$. Unlike in the repulsive case
of Sec.~\ref{rep} where the corresponding solutions diverge rapidly, the attractive solutions are always oscillatory. The condition that $|\psi(0)| >> 1$ implies that the  nonlinearity in eq.~(\ref{eq_a1}) takes very small values.
This is because the chosen large values for
$\psi(0)$ scale as $1/\sqrt{|g|}$, where $g$ is the nonlinear amplitude
of $|\psi|^2$ in the normalization given by Mallory \& Van Gorder \cite{mal13,mal14a,mal14b,mal15} for eq.~(\ref{eq_a1}).
This is a surprising result as it indicates that exotic bright solitons appear only for small perturbations (of order $g|\psi|^2$) in the {Schr{{\"o}}dinger} equation,
while strong perturbations with $0 < |\psi(0)| < 1/\sqrt{2}$ always lead to well-behaved oscillatory solutions about $\psi = \psi_+$ or $\psi = \psi_-$ (as the red curve in Fig.~\ref{fig2}).

\section{Discussion}\label{sum}

We have presented an analysis and a classification of the oscillatory properties of the solutions of the cubic nonlinear {Schr{{\"o}}dinger} equation \cite{mal13,mal14a,mal14b,mal15,lif80}.
The analysis makes use of a procedure that was originally described in \cite{chr16a,kat16}
for second-order linear homogeneous differential equations.
It turns out that the same procedure is also valid for nonlinear homogeneous equations, provided that they possess at least one trivial solution 
that may serve as a baseline for oscillations. This requirement is oversatisfied
by the CNLS equations in one, two, and three spatial dimensions as they possess 3 different trivial solutions. The presence of so many trivial solutions is the driver
for all the oscillatory properties seen in the solutions of the boundary-value problem in both the repulsive and the attractive case (Figs.~\ref{fig1} and~\ref{fig2}, respectively); 
and the solutions in these two cases differ only because the trivial solutions 
interchange their roles from repelling to attracting the nontrivial solutions
and vice versa. 

We carried out our analysis simultaneously for all CNLS equations because the applicable criteria for oscillatory behavior reduce asymptotically to the simple inequality~(\ref{crit_3})
that effectively requires a positive coefficient in front of the non-derivative $\psi$-terms in
eqs.~(\ref{eq_r1}) and~(\ref{eq_a1}). We have confirmed numerically that
the oscillation criterion derived in the linear case \cite{chr16a,kat16} carries over to the CNLS equations
as well (Figs.~\ref{fig1} and~\ref{fig2}). This is the direct result of the behavior of the inertial terms in the nonlinear
1+1, 2+1, and 3+1 BEC cases (0, $\psi'/x$, and $2\psi'/x$, respectively).

We have also found evidence for asymmetric behavior between the repulsive and the attractive CNLS free-particle solutions beyond of the known difference in the velocities of the two types of solitons \cite{pet08}. The attractive stationary case supports a new  physical oscillatory solution that appears when a boundary condition 
with $|\psi(0)| >> 1$ is used for the bright wavefunction (Fig.~\ref{fig3}). 
In the corresponding repulsive stationary case, 
the dark wavefunctions are all diverging steeply and they do not appear to be of
physical interest \cite{mal14b}. The oscillatory features and the divergent behavior discussed in this work are the result of the nonlinear terms in the free-particle CNLS equations; and they remain intact in cases where various external potentials are used \cite{mal13,mal14a,mal14b,mal15} to model traps for various BECs. In both cases, the boundary condition
that  $|\psi(0)| >> 1$ implies that the nonlinearities in the equations
are very small, so these solitons appear only for small perturbations
(of order $g|\psi|^2$, where $|g|<< 1$)
in the {Schr{{\"o}}dinger} equation. This is a surprising result.
The bright (exotic) oscillatory stationary solutions of Fig.~\ref{fig3} 
may actually be identifiable in real
BECs because they exhibit a strong elongation in their radial wavelength
(by factors of order $\sim$2; Sec.~\ref{att}) at intermediate radii.

Taken all together, our results lead to another interesting conclusion:
the presence or the absence of trivial solutions in differential equations
of the second order is an important qualifier of the properties of the
solutions of the physical Cauchy problem; thus, they should not be ignored,
as their name signifies. Lately, we have come to call them {\it intrinsic} solutions \cite{chr16b}
because when the differential equations admit such solutions, they do so
with no regard to any boundary or initial conditions that may be imposed
externally by the Cauchy problem.

\subsection*{Acknowledgments}
During this research project, DMC was supported by the University of
Massachusetts Lowell whereas QDK was on a sabbatical visit and was fully supported by the Jordan University of Science and Technology.



\address{Department of Mathematics and Statistics, \\
Jordan University of Science and Technology, \\
Irbid, Jordan 22110 \\
\email{qutaibeh@yahoo.com} \\
\\
Department of Mathematical Sciences, \\
University of Massachusetts Lowell, \\
Lowell, MA 01854, USA \\
\email{dimitris\_christodoulou@uml.edu} \\
}

\end{document}